\documentclass[aps,showpacs,prb,twocolumn]{revtex4}
\usepackage{times,xspace}
\usepackage{amsbsy,amssymb,amsmath,bm}
\usepackage{graphicx,color,epsfig,subfigure,rotate}

\begin{document}

%\draft
%\preprint{}
\title{Interaction of a discrete breather with a lattice junction}
\author{Ioana Bena$^a$, Avadh Saxena$^{a,b}$, and J. M. Sancho$^a$
%\address{} 
\\$^a$Department d'Estructura i Constituents de la Mat\`eria,
Universitat de Barcelona, 08028 Barcelona, Spain
\\$^b$Theoretical Division, Los Alamos National Laboratory,
Los Alamos, New Mexico 87545 }  

\date{\today}

\begin{abstract}

{We study the scattering of a moving discrete breather (DB) on a
junction in a Fermi-Pasta-Ulam (FPU) chain consisting of two
segments with different masses of the particles.  We consider four
distinct cases: (i) a light-heavy (abrupt) junction in which the
DB impinges on the junction from the segment with lighter mass,
(ii) a heavy-light junction, (iii) an up mass-ramp in which the
mass in the heavier segment increases continuously as one moves
away from the junction point, and (iv) a down mass-ramp.  
Depending on the mass difference and DB characteristics 
(frequency and velocity), the DB can either reflect from, or
transmit through, or get trapped at the junction or on the ramp.
For the heavy-light junction, the DB can even split at the
junction into a reflected and a transmitted DB.  The latter is
found to subsequently split into two or more DBs.  For the down
mass-ramp the DB gets accelerated in several stages, with
accompanying radiation (phonons).  These results are rationalized   
by calculating the Peierls-Nabarro barrier for the various cases.
We also point out implications of our results in realistic
situations such as electron-phonon coupled chains. }               
\end{abstract}  


\maketitle 

\section{Introduction} 
\label{introduction}

Static discrete breathers (DB) are time-periodic, persistent, intrinsic 
localized {\it exact} modes in nonlinear lattices. Rigorous proofs of 
their existence have been obtained and systematic studies of their 
properties were carried out using various (approximate) complementary 
approaches, see e.g.  [\onlinecite{Flach1998}] for an overview.  
In contrast, as first noticed in numerical investigations and then 
justified theoretically, moving DBs exist as {\it approximate} solutions 
in nonlinear lattices, both Hamiltonian and non-Hamiltonian (with
dissipation and periodic forcing). These solutions are known to be 
rather stable (i.e., long-lived) and have been an object of constant 
investigation during the last decade, see  [\onlinecite{moving}] for a 
non-exhaustive list.

Different physical systems in which there are realizations of (moving) 
DBs include conjugated polymers [\onlinecite{kress,avn2}], 
charge-density-wave 
materials (e.g., metal-halogen electronic chains [\onlinecite{avn3}]), 
Josephson ladders [\onlinecite{avn4}], coupled electron-vibron lattice systems 
[\onlinecite{hennig}] and spin chains [\onlinecite{spin}].  Sputtering on
crystal surfaces and damage tracks in certain mica minerals have also been 
attributed to moving breathers [\onlinecite{mica}].
Experimentally, breathers 
have been probed by ultrafast resonance Raman [\onlinecite{avn3}] 
and inelastic 
neutron scattering [\onlinecite{fillaux}] among other techniques.

In a recent series of papers [\onlinecite{impurity}], the problem of the
interaction of a moving DB with an impurity  was addressed in the case 
of a lattice with nonlinear on-site potential and harmonic first-neighbor 
coupling. As it was shown, this interaction can lead to reflection, 
transmission or trapping of the DB at the impurity, depending on the 
initial velocity, amplitude and phase of the DB, as well as on the 
`strength' and spatial extent of the impurity.  

Our objective here is to investigate the scattering of a DB at a 
junction in a (nonlinear) FPU chain consisting of two segments that are 
``slightly different"--i.e., for instance, with different interaction 
parameters or with different masses of the particles in the two segments. 
The reason for choosing the FPU chain is simple.  It has historically 
provided a testbed for exploring novel nonlinear phenomena in discrete 
systems.  In addition, it is one of the simplest nonlinear (polynomial) 
potentials amenable to some analytical calculations.   

Our preliminary numerical simulations indicate that these two types of 
problems are not qualitatively very different.  Therefore, here we will 
concentrate exclusively on the second type of configuration, the one 
with slightly different masses on the two sides of the chain. A 
physical 
realization of this configuration could be in 
low-dimensional electronic materials with 
different electron-phonon coupling or two segments with different 
isotopes (e.g., carbon isotopes in conjugated polymers 
[\onlinecite{kress,avn2}] and platinum 
isotopes in metal-halogen chains [\onlinecite{avn3}]), 
Josephson-Junction arrays 
[\onlinecite{avn4}]
with dissimilar interaction strengths, optical fibers with two 
different refractive indices
[\onlinecite{agrawal}], etc.  We note that the scattering of 
Toda solitons at a mass interface was studied previously 
[\onlinecite{mertens1}].  
To the best of our knowledge, the scattering of a DB at such an interface 
has not yet been investigated.

The paper is organized as follows.  In Sec. II we present the details of 
the FPU model in a homogeneous chain, an estimate of the Peierls-Nabarro 
barrier for moving DBs, and finally, some details on the numerical 
initialization of a moving DB.  Section III contains results for 
both the light-heavy and the heavy-light mass junction, where we 
elaborate on the reflection and transmission (and eventually on the 
splitting) of the DB.  Interaction of the DB with both the up mass-ramp 
and the down mass-ramp is discussed in Sec. IV, where we explore DB 
reflection (with eventual trapping) and acceleration (with eventual 
splitting). In Sec. V we summarize our main findings and enumerate some 
of the open questions.  Details of the Peierls-Nabarro barrier 
calculation, using a new perturbative technique, for the various 
homogeneous and inhomogeneous cases are relegated to an Appendix.  

\section{The model}
\label{model} 

\subsection{The FPU model} 
\label{FPU}

The FPU model represents a one-dimensional (1D) chain of particles with 
no on-site potential (i.e., an acoustic chain), with the Hamiltonian
\begin{eqnarray}
H&=&\sum_{n}\left[\frac{m\dot{x}_n^{2}}{2}+\frac{\alpha}{2}(x_{n+1}-x_n-a)^2
\right.\nonumber\\
&+&
\left.\frac{\beta}{4}(x_{n+1}-x_n-a)^4\right]\,,
\label{FPUHamiltonian}
\end{eqnarray}
where $\alpha$ and $\beta$ denote, respectively, 
the strengths of the linear and nonlinear nearest-neighbor interactions;
$a$ represents the lattice constant (i.e., the equilibrium distance
$a=x_{n}^{eq}-x_{n-1}^{eq}$ between neighboring sites), 
and $m$ is the mass of the particles. 
For simplicity, 
all these quantities (and those we will introduce later)
are expressed in 
{\it dimensionless units}.
The
corresponding equation of motion for a generic particle is:
\begin{eqnarray}
m\ddot{x}_{n}&=&\alpha\,(x_{n+1}+x_{n-1}-2x_{n})\,+\,\beta\,
[(x_{n}-x_{n-1}-a)^3 \nonumber \\ 
&-&(x_{n+1}-x_{n}-a)^3]\,.
\end{eqnarray}
In terms of the elongations $u_n = x_n - x_n^{eq}$, it becomes:
\begin{eqnarray}
m\ddot{u}_{n}&=&\alpha\,(u_{n+1}+u_{n-1}-2u_{n})\,+\,\beta\,
[(u_{n}-u_{n-1})^3\nonumber\\
&-&(u_{n+1}-u_{n})^3]\,,
\label{elongations}
\end{eqnarray}
or, by introducing the {\it relative} elongations of neighboring sites
\begin{equation}
\tau_n=(x_{n}-x_{n}^{eq})-(x_{n-1}-x_{n-1}^{eq})=x_{n}-x_{n-1}-a\,,
\end{equation}
\begin{equation}
m\ddot{\tau}_{n}=\alpha\,(\tau_{n+1}+\tau_{n-1}-2\tau_{n})\,+\,\beta\,
(\tau_{n+1}^3+\tau_{n-1}^3-2\tau_{n}^3)\,.
\label{relativeelongations}
\end{equation}
As it was shown (see, for example  [\onlinecite{Flach1998}] for an 
overview, and references therein, and  [\onlinecite{James2001}]), 
the FPU lattice admits DB-like solutions (stationary, localized, 
time-periodic modes) with periods $T_{DB}$ that are smaller than the 
minimum period of the phonon spectrum, i.e.,
\begin{equation}
T_{DB} < \pi \,\sqrt{m/\alpha}\,. 
\end{equation}
Also, as shown, for example in  [\onlinecite{Page1990}] and  
[\onlinecite{Sandusky1992}], the most localized of these modes are an {\it 
odd-type} mode with an ``approximate'' pattern of the amplitudes of 
the elongations $u_n$ of the form: $A_{odd}(...,0,-1/2,1,1/2,0,...)$, 
and an {\it even-type} mode [\onlinecite{footnote1}]~: $A_{even}(...,0,-1,1,0,...)$.  
The amplitudes $A$ are determined by the interaction constants $\alpha$
and $\beta$ and by DB's frequency $\omega_{DB}=2\pi/T_{DB}$.  For 
given interaction constants, the $A$'s decrease with increasing $T_{DB}$.  
On the contrary, keeping $T_{DB}$ fixed and decreasing the interaction
constants generally leads to an increase in the amplitudes $A$.  
By ``approximate" 
above we mean that (as seen in  [\onlinecite{Sandusky1992}]) these 
patterns are exact only for a pure even-order anharmonic lattice in the 
limit of increasing order of anharmonicity. Nevertheless, only minor 
corrections are needed in order to make these patterns `more precise' 
solutions of the FPU lattice, their symmetry being preserved. Mainly, 
these corrections refer to the fact that the DB can extend over more 
than three, and two sites, respectively, for odd and even modes.  
Although, in theory, a DB has an infinite extension [with an exponential 
decay of the amplitude of the relative elongation as one moves far away 
from the center (maximum amplitude sites) of the DB], in practice, 
however, one can restrict the analysis to five, and four sites, 
respectively, for the two types of modes of the DB mentioned above. 

To evaluate the relative elongations for the two configurations, 
and their corresponding energies, we introduced a simple perturbative 
technique that uses the ratio between the square of the maximum phonon 
frequency and the square of the DB frequency as the perturbation parameter:
\begin{equation}
\varepsilon = \frac{4 \alpha}{m \, \omega_{DB}^2}\,,
\label{epsilon}
\end{equation} 
combined with a rotating wave approximation, RWA (see, e.g.,  
[\onlinecite{Flach1998}] and references therein).  The results of our 
calculations, presented below and, in more detail, in the Appendix, 
can be compared with the numerical results of Green's function method 
(that is also based on RWA).  For example, for 
the even-symmetry mode, our calculations [up to ${\cal O}(\varepsilon ^2 )$]
agree generally up to an error of no more than $4 \%$ with the results of
 [\onlinecite{Sievers1988}] obtained with Green's function method.  The 
error in evaluating the configurations (as compared with the results of 
the {\it exact} numerical method of the analytical continuation from the 
anticontinuous limit [\onlinecite{Aubry}]) is essentially connected with 
the limitations of RWA, and therefore becomes progressively smaller for 
`heavier' DBs, i.e., DBs that are progressively further away (in frequency)
from the 
phonon band limit.

The primary ingredients of the analytic method are the ansatz concerning 
the temporal evolution of particles' elongations:
\begin{equation} 
u_n(t) = A \xi_n \cos(\omega_{DB} t) \,,
\end{equation}
(where $A$ and $\xi_n$ are the amplitude and the shape function, 
respectively), together with the RWA that entails neglecting 
higher-frequency harmonics [i.e., $\cos^3(\omega_{DB} t) \approx (3/4) 
\cos(\omega_{DB} t)$].  Including these elements in Eq. (\ref{elongations}), 
one obtains an infinite set of nonlinear coupled equations for the shape 
function:
\begin{eqnarray}
m\omega_{DB}^2 \xi_n &=& \alpha (2 \xi_n -\xi_{n+1} - \xi_{n-1}) \nonumber \\ 
&+& \frac{3}{4} \beta A^2[(\xi_n -\xi_{n+1})^3+(\xi_n -\xi_{n-1})^3]\,. 
\label{xi}
\end{eqnarray}
Or, in terms of the {\it relative} elongations:
\begin{equation} 
\tau_n(t) = A \zeta_n \cos(\omega_{DB} t)\,,\,\,\, \zeta_n=\xi_n-\xi_{n-1}\,,
\end{equation}
\begin{eqnarray}
m\omega_{DB}^2 \zeta_n &=& \alpha (2 \zeta_n -\zeta_{n+1} - \zeta_{n-1}) 
\nonumber \\  
&+& \frac{3}{4} \beta A^2(2 {\zeta_n} ^3-{\zeta_{n+1}}^3 -{\zeta_{n-1}}^3)\,\,.
\end{eqnarray}
Next, we consider the following expansion of the shape function in terms 
of the small parameter $\varepsilon$, see Eq. (\ref{epsilon}):
\begin{equation}
\xi_n = \xi_n^{(0)} + \varepsilon \xi_n^{(1)} + \varepsilon^2 \xi_n^{(2)} +...
\label{seriesxin}
\end{equation}
and then proceed through the usual steps of a perturbative calculation.
For details of these calculations, refer to the Appendix.  

\subsubsection{The Peierls-Nabarro barrier for the homogeneous FPU chain}
\label{PNBhomogeneous}

\begin{figure}
\subfigure[homogeneous chain]{
\begin{minipage}{0.5\textwidth}
{
\psfig{figure=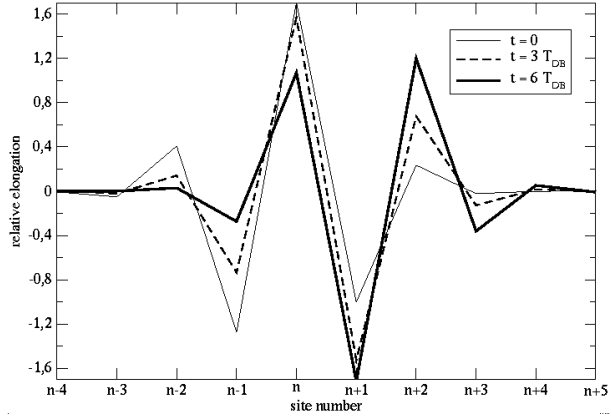,width=8cm}}
\label{figure1a}
\end{minipage}
}\vspace{0.5cm}\\
\subfigure[light-heavy junction, $m_B=1.002$: transmission of the DB]{
\begin{minipage}{0.5\textwidth}
{
\psfig{figure=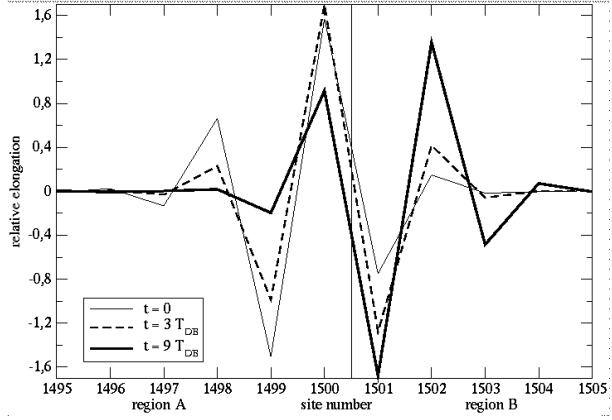,width=8cm}}
\label{figure1b}
\end{minipage}
}\vspace{0.5cm}\\
\subfigure[light-heavy junction, $m_B=1.04$: reflection of the DB]{
\begin{minipage}{0.5\textwidth}
{
\psfig{figure=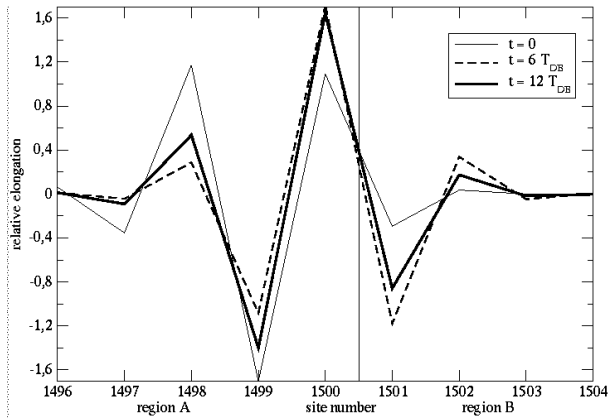,width=8cm}}
\label{figure1c}
\end{minipage}
}
\caption{Temporal evolution of a DB's configuration 
in three different situations. 
Notice the alternation between odd and even- type
configurations.
$T_{DB}=2.1$. }
\label{figure1}
\end{figure}

\begin{figure}
{
{\psfig{figure=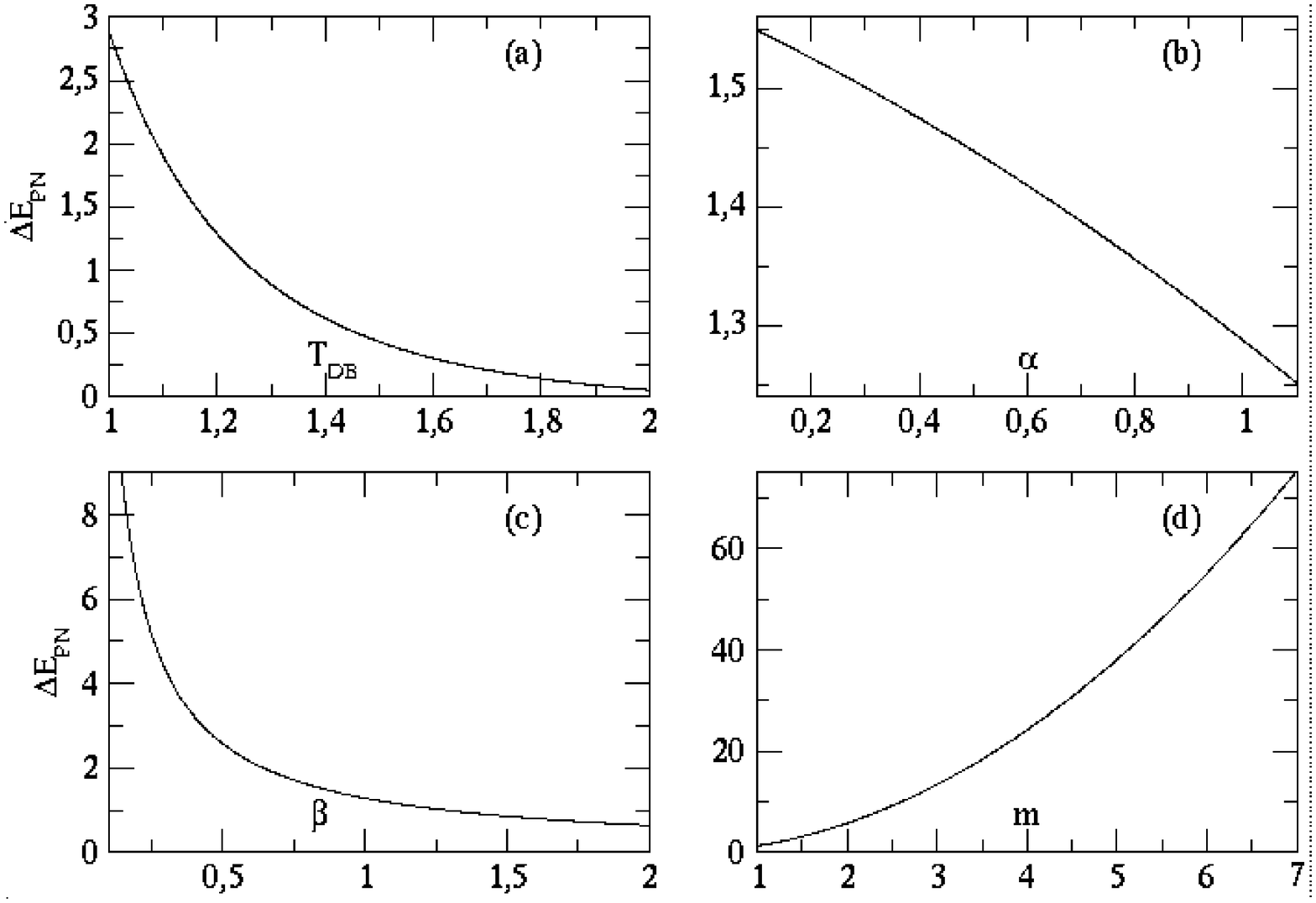,width=8cm}}}
\caption{Variation of the Peierls-Nabarro barrier as a function of 
breather time period ($T_{DB}$), FPU chain parameters ($\alpha$, $\beta$) 
and mass ($m$).  (a) $\alpha=1$, $\beta=1$, $m=1$. (b) $T_{DB}=1.2$, 
$\beta=1$, $m=1$.  (c) $\alpha=1$, $T_{DB}=1.2$, $m=1$. (d) $\alpha=1$, 
$\beta=1$, $T_{DB}=1.2$.}
\label{figure2}
\end{figure}

As illustrated in Fig.~\ref{figure1a} on an actual example, 
the DB translates from one lattice site to another by 
continuously deforming its shape, alternately, 
between an odd-type of 
configuration and an even-type one.  Therefore, 
in a discrete lattice there is an energy cost associated with moving 
a nonlinear localized mode by a lattice constant--this represents
the so-called 
Peierls-Nabarro barrier (PNB), see [\onlinecite{PNBARRIER}].  
It can be estimated by calculating the 
energy difference between even- and odd-type configurations.  The 
results presented in the Appendix allow us to evaluate the PNB in an 
homogeneous chain (i.e., all particles with same $\alpha$, $\beta$, and $m$):
\begin{eqnarray}
\Delta E_{PN}^{h} &=& E_{odd}^{h}-E_{even}^{h}=m \omega_{DB}^2
\left(\frac{\alpha}{\beta}\right)\left[{0.00836}\,{\varepsilon}^{-1} \right.
\nonumber \\  
&&-\,\left.0.00765 -0.01827\, \varepsilon +
{\cal O}(\varepsilon^2)\right]\,,
\label{PNBh}
\end{eqnarray}
where the superscript $h$ refers to the homogeneous case.
As expected, it is a very small energy barrier (as compared with the one
typically found in some optical chains, i.e., chains with on-site nonlinear
potential, see [\onlinecite{PNBARRIER}]); for example, for a very heavy DB,
$\Delta E_{PN}^{h}/E_{odd}^{h} \sim 2.1 \%$ only! This  explains the 
well-known fact that it is rather easy to create mobile DBs in an FPU chain, 
and also why in the first-order approximation in [\onlinecite{McKay2002}] 
this barrier was found to be zero. In Fig.~\ref{figure2} we represent the 
dependence of the barrier on various parameters: (i) DB's period $T_{DB}$ 
(as expected, also see below the discussion on the generation of moving
DBs, the PNB is larger for higher-frequency DBs; in the first order of the
perturbative expansion, PNB varies as $1/T_{DB}^4$).  
(ii) $\alpha$ and (iii) $\beta$. 
At the first order in the perturbational expansion, the PNB does not depend
on $\alpha$, but only on $1/\beta$, i.e., it decreases with increasing 
nonlinearity.  This feature can be easily understood if one views the role 
of the nonlinearity as reducing particles' excursions around equilibrium, 
and therefore reducing the differences between the odd- and even-parity 
configurations, i.e., the PNB.  (iv) Finally, on $m$ (note that, in the 
first order of the perturbative expansion, the PNB  varies as $m^2$).

\subsection{DB generation and initialization}
\label{DBgeneration}
For simulation purposes,  
the static DBs were generated numerically in the homogeneous FPU chain  
using the extremely fast algebraic method recently introduced by Tsironis 
[\onlinecite{Tsironis2002}].  As shown in [\onlinecite{Tsironis2002}], 
this method, although approximate, is generally more accurate than the 
RWA and agrees with the exact results of the anticontinuous limit method 
(which requires much longer computational times, see [\onlinecite{Aubry}]) 
typically to $1\%$ or even better.

In order to move these breathers, we used a simple approximation of the 
systematic pinning mode excitation method of Chen et al. 
[\onlinecite{Chen1996}]. 
Namely, we `kick' initially the DB by assigning to the points of the 
lattice initial relative velocities that are a fraction $\lambda$ (the 
``kicking coefficient") of the gradient of $|\tau_n|$, i.e.,
\begin{equation}
\dot{\tau_n}(t=0)=\lambda\,(|\tau_{n+1}|-|\tau_{n-1}|)/2 \, \,.
\label{kick}
\end{equation}
Note that this method is not so different from the `more empirical' 
methods used in [\onlinecite{Sandusky1992}] to obtain 
moving DBs.  We notice that, when starting to move, the DB first 
loses, through phonon radiation in the lattice, a large part 
of the kinetic energy we assigned to it when kicking. 
The rest of the received energy is used to overcome the Peierls-Nabarro 
barrier, and, as already mentioned, 
the DB moves from one lattice site to another by a continuous alternation
between odd- and even- type configurations.

This alternation between the two types of configurations for a moving 
DB can be noticed when inspecting the temporal evolution of the potential 
(or kinetic) energy of the DB. Indeed, the envelope of the temporal 
oscillations of DB's potential (kinetic) energy presents a series of 
periodically alternating relative maxima and minima, indicating the 
alternation between these configurations.  The period between two such 
successive maxima (or minima) gives a rough estimate of the time needed 
by the DB to move from one site to another. But no more than a ``rough 
estimate", because (i) the real time a DB takes for this movement does 
not bear a commensurability relation with $T_{DB}$, and (ii) the structure 
of the envelope is more complex, due to the presence of other ``secondary"
frequencies of the DB (see [\onlinecite{Flach1998}], and the 
discussion in Sec. V), and (iii) there are some ``imperfections" in this 
periodic behavior of the envelope, that are connected to the existence of 
a rather irregular time dependence of the relative phases of two 
neighboring sites (already mentioned in [\onlinecite{Sandusky1992}]). 
Probably, this is ultimately related to the non-exact character of a 
moving DB as a solution of the Hamiltonian lattice.  Note also that a 
moving DB  constantly loses energy while moving through the lattice, 
although at a very small `dissipation rate'. For example, as also shown 
in [\onlinecite{Ibanes2002}], for a DB moving in a uniform lattice, this 
energy decrease, if fitted to an exponential, corresponds to a decay 
rate on the order of $\sim 10^{-6}$/unit time. This rate is higher for 
the faster DBs.  Also, as explained in detail by the same authors, the 
analysis of the temporal behavior (and, in particular, of the extremal 
points) of the kinetic and potential energy allows one to evaluate the 
translational energy of a moving DB, which was found to be at most $1 \%$ 
of the total energy of the DB. Not surprisingly, this value is of the 
same order of magnitude as the Peierls-Nabarro barrier.

Returning to the kicking method for moving a DB, we make several 
other remarks. First that, as previously noticed (see, e.g., 
[\onlinecite{Dauxois1993}]), the `light' DB's (i.e., those that 
are relatively not too far in frequency above the phonon band limit) 
are definitely 
easier to move than the `heavy' DBs (which are, by comparison, much 
more localized and therefore much more sensitive to the discreteness 
of the lattice). In terms of the initial kick, this means that the 
minimum value of the kicking coefficient, $\lambda$,
for which one gets an essentially regular 
motion of the DB [\onlinecite{footnote2}]
is larger for `heavier' DBs.  Also, one notices that, in general, the 
velocity of the moving DB obtained through this kicking method seems 
first to increase with increasing $\lambda$, and after that it reaches 
a certain `saturation value', i.e., it does no longer increase with 
$\lambda$, but keeps a constant value. This leads to a rather narrow 
window of the possible values of the DB velocities, which is somewhere 
around a tenth of the phonon's velocity (for example, for $\alpha 
=\beta=1$, a lattice constant $a=10$, and for a DB of period $T_{DB}=2.1$, 
the values of the velocities belong to a window of $\approx [0.35, 1.25]$; 
note that the sound velocity corresponds to $10$ in the dimensionless units).  
In the simulations we used a 
chain with the first and last point held fixed (i.e., fixed boundary 
conditions; this should not raise conceptual problems, as such points 
correspond to $m \rightarrow \infty$).  Also, we tried to avoid the 
interference between the observed phenomena and the phonons that reflect 
on these fixed edges--and for this purpose we generally used sufficiently 
long chains (so that the reflected phonons do not come back to the 
interesting central regions during the observation period).

\section{Interaction of a DB with a junction}
\label{junction}
 
We now address the main problem in this paper. Consider the junction 
between two semi-infinite FPU chains (let us call them ``A" and
``B", respectively, with the corresponding subscripts for their 
characteristic parameters). We fix the parameters of the A chain 
$\alpha_A$, $\beta_A$, and $m_A$. For the B chain, we will fix the 
interaction parameters identical to those of the A chain,
\begin{equation}
\alpha_B =\alpha_A\,\,,\,\,\,\,\,\,\beta_B=\beta_A\,,
\end{equation}  
and vary the mass of the particles, $m_B$.  Note that in all the 
numerical simulations we set $\alpha_A=\beta_A=1$, as well as $m_A=~1$.

A DB is generated in the A part of the chain and is sent to the 
junction with the B part. Depending on the difference between the 
masses of the particles in the two parts of the chain, the DB 
exhibits different behaviors at  the junction.

\subsection{The Peierls-Nabarro barrier at  the junction}
\label{PNBjunction}

In order to understand and predict the behavior of a moving DB at 
such a junction, the first step is to study the change 
in the Peierls-Nabarro barrier at the junction. Namely, 
to determine what would be the equivalent of the odd- and even- type 
configurations at the junction, and what would be the corresponding 
difference in the configurational energy.  
Note that in the case of an inhomogeneous chain the PNB 
is defined as the difference 
between the {\it global} maximum and the 
{\it global} minimum of the configurational 
energy.  We consider the case when 
$\delta = (m_B-m_A)/m_A$ is a small quantity, that we use 
as a perturbation parameter for evaluating the changes in the odd 
and even configurations at  the junction.  Thus, we 
consider a ``double" perturbation expansion of particles' 
envelope function:
\begin{eqnarray}
\xi_n &=&(\xi_{n,0}^{(0)} + \varepsilon_A \xi_{n,0}^{(1)} + 
\varepsilon_A^2 \xi_{n,0}^{(2)} +...)\nonumber\\
&+& \delta \,\,(\xi_{n,1}^{(0)} + \varepsilon_A \xi_{n,1}^{(1)} + 
\varepsilon_A^2 \xi_{n,1}^{(2)} +...)\nonumber\\
&+& \delta ^2\,\,(\xi_{n,2}^{(0)} + \varepsilon_A \xi_{n,2}^{(1)} + 
\varepsilon_A^2 \xi_{n,2}^{(2)} +...)+...\,,
\label{doubleseries}
\end{eqnarray}
where $\varepsilon_A$ is evaluated with respect to the parameters of
the A chain, i.e.,
\begin{equation}
\varepsilon_A=\frac{4 \alpha_A}{m_A \, \omega_{DB}^2}\,,
\end{equation}
Corresponding to the different configurations it has to exhibit in order 
to traverse the junction, the DB encounters three new energy barriers  
(refer to the Appendix for more details). These are, in the 
order of their appearance as the DB moves through the junction: 
\begin{eqnarray}
\Delta E_{PN}^{j(I)}&=&\Delta E_{PN}^{h(A)}+
m_A\omega_{DB}^2 \left(\frac{\alpha_A}{\beta_A}\right)\times\nonumber\\
&&\hspace{-2cm}\times\left[\delta
\left(0.13793\, \varepsilon_A^{-1}-0.01573 + 0.0056\, \varepsilon_A\right)\right.
\nonumber\\
&&\hspace{-2cm}\left.+\delta^2
\left(0.49933 \,\varepsilon_A^{-1}+0.21657 -0.69346\, \varepsilon_A\right)\right]. 
\end{eqnarray} 
This energy barrier  corresponds to the difference between the energy of 
the odd-type configuration I in the Appendix--for which the site of maximum 
elongation is the last site in the part A of the chain--and the energy of 
the even-type configuration in the homogeneous A chain.  
\begin{eqnarray}
\Delta E_{PN}^{j(II)}&=&\Delta E_{PN}^{h(A)}+
m_A\omega_{DB}^2 \left(\frac{\alpha_A}{\beta_A}\right)\times\nonumber\\
&&\hspace{-2cm}\times\left[\delta
\left(0.64113 \,\varepsilon_A^{-1}-0.14880 -0.00526 \,\varepsilon_A\right)\right.
\nonumber\\
&&\hspace{-2cm}\left.+\delta^2
\left(0.75093\, \varepsilon_A^{-1}+0.21657 -0.68820 \,\varepsilon_A\right)\right]. 
\label{barrierII}
\end{eqnarray} 
This energy barrier  corresponds to the difference between the energy of the 
odd-type configuration III in the Appendix--for which the site of maximum 
elongation is now the first site in the part B of the chain--and the energy 
of the even-type configuration in the homogeneous A chain).  The last PNB 
barrier is associated with the difference between the energy of the odd-type 
configuration in the homogeneous B chain and the even-type configuration in 
the homogeneous A chain:
\begin{eqnarray}
\Delta E_{PN}^{j(III)}&=&\Delta
E_{PN}^{h(A)}+m_A\omega_{DB}^2
\left(\frac{\alpha_A}{\beta_A}\right) \nonumber\\
&&\hspace{-2cm}\times\left[\delta
\left(0.77906 \,\varepsilon_A^{-1}-0.16470\right)\right.
\hspace{-0.1cm}\left.+\delta^2  
\left(0.38953\, \varepsilon_A^{-1}\right)\right]\,.
\end{eqnarray}
Here $\Delta E_{PN}^{h(A)}$ denotes the Peierls-Nabarro barrier in the 
homogeneous A chain, and $j$ refers to the junction.

Note that for a heavy-light junction, i.e., for $\delta <0$, these barriers
are smaller than the PNB barrier in the homogeneous A part of the chain 
and therefore a DB that moves smoothly in region A will have no `energetic 
difficulties' to enter  region B.  On the contrary, for a light-heavy 
junction, i.e., for  $\delta > 0$, these barriers are larger than the PNB 
in part A of the chain and one sees that, at the dominant orders in 
$\varepsilon_A$ and $\delta$, they increase in succession.  Therefore, 
there appears the possibility that a DB that arrives at such a junction 
cannot overcome either the first, or the second , or the third barrier. 
The presence of these barriers is confirmed by numerical simulations 
(through fine tuning of $m_B$).
  
\subsection{The light-heavy junction}
\label{lightheavy}

We first present the generic results of our simulations for this case.

\begin{figure} 
{\psfig{figure=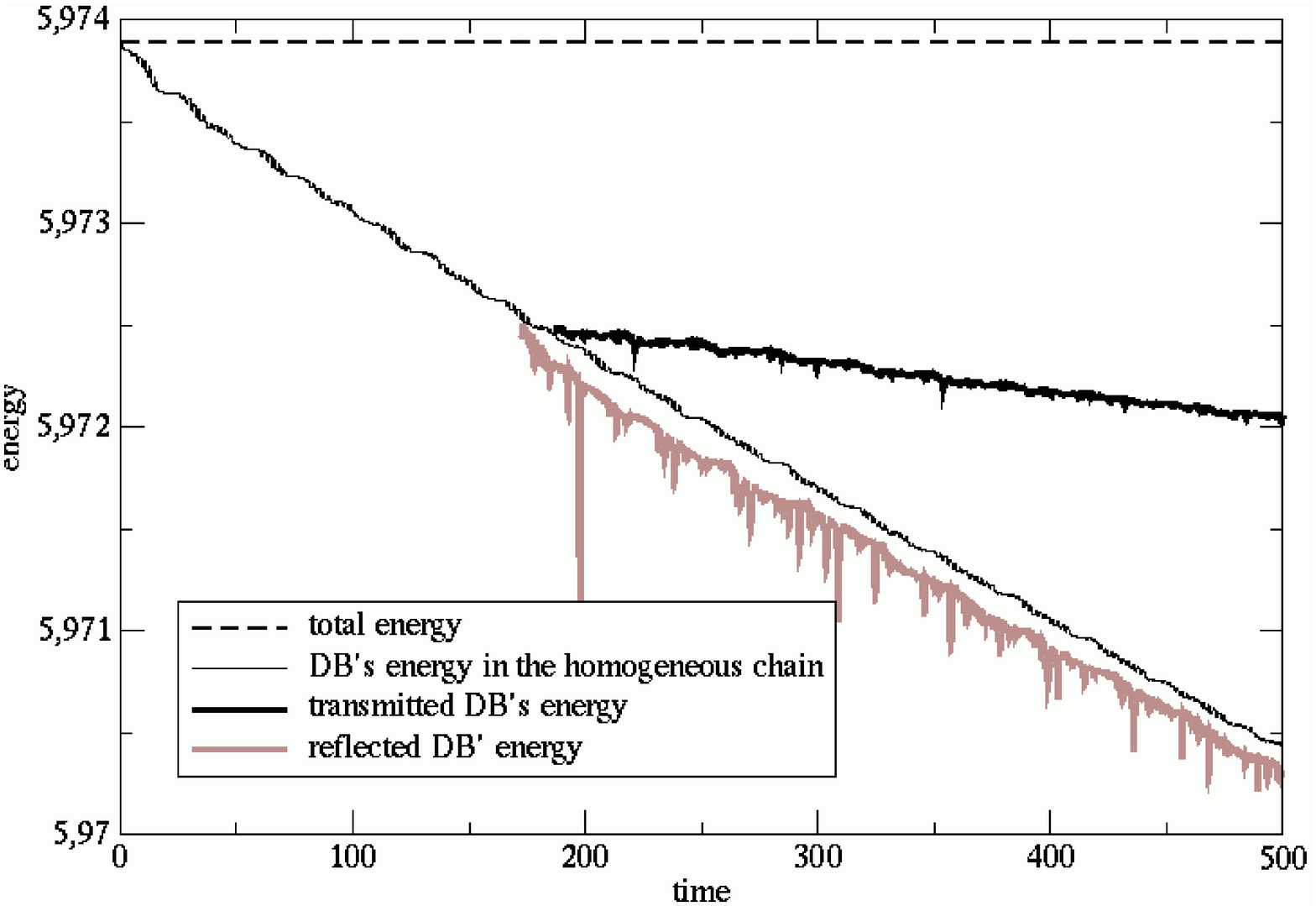,width=8cm}}
\caption{Energy of the DB for the three cases in Fig. \ref{figure1}.}
\label{figure3}
\end{figure}

(a) A DB can continue its movement into region B.  Its frequency is not
(detectably) modified.  The DB keeps on losing energy in region A as 
well as in region B, but at a smaller rate in region B, see Fig.~\ref{figure3}.  This might be connected to the fact that (given 
that it keeps essentially the same frequency) the DB is further away
from the phonon band limit in region B than in region A.  Also, its velocity 
in region B is smaller than in region A. This is related to the fact 
that a part of the ``extra" energy that in region A corresponded to 
its movement as a whole (with a velocity $v_A$) is now used 
for the new,  higher mean configurational  energy, and also
to overcome  the correspondingly
 higher Peierls-Nabarro barrier in region B.
Therefore, the  ``extra"  kinetic energy, 
and correspondingly the velocity $v_B$
in  region B are smaller than in region A.

(b) The DB can reflect at the junction and return to the 
region A.  Its frequency and energy (see Fig.~\ref{figure3}) are not 
sensitively modified by this reflection, and neither its velocity (that 
only changes sign). 

These observations can be explained qualitatively on the basis of the 
results  presented above for the PNB that a DB 
(that keeps a constant period $T_{DB}$)  has to overcome in order to 
continue its movement in region B. 
The main conclusion is that for a DB arriving at the junction, there 
exists a critical value of the mass $m_B=m_{crit}^j$ above which the DB
cannot  penetrate in region B and is reflected to region A.  This 
critical value depends on the frequency of the DB, namely it increases 
with decreasing $T_{DB}$ (i.e., it is larger for heavier DBs). However, 
it also depends on the velocity the DB has in region A: it increases 
with increasing $v_A$ (i.e., a more rapid breather needs a larger 
mass $m_{crit}^j$ to be reflected than a slower DB of the same frequency). 
This can be readily understood: a more rapid DB in region A has more
``extra" energy (above the Peierls-Nabarro barrier $\Delta E_{PN}^{h(A)}$) 
than a slower one.  Therefore, it may use this energy to overcome the 
Peierls-Nabarro barrier at the junction and to penetrate 
in region B, while a slower DB cannot overcome the junction barrier.

\begin{figure}
\subfigure[homogeneous chain]{
\begin{minipage}{0.5\textwidth}
{
\psfig{figure=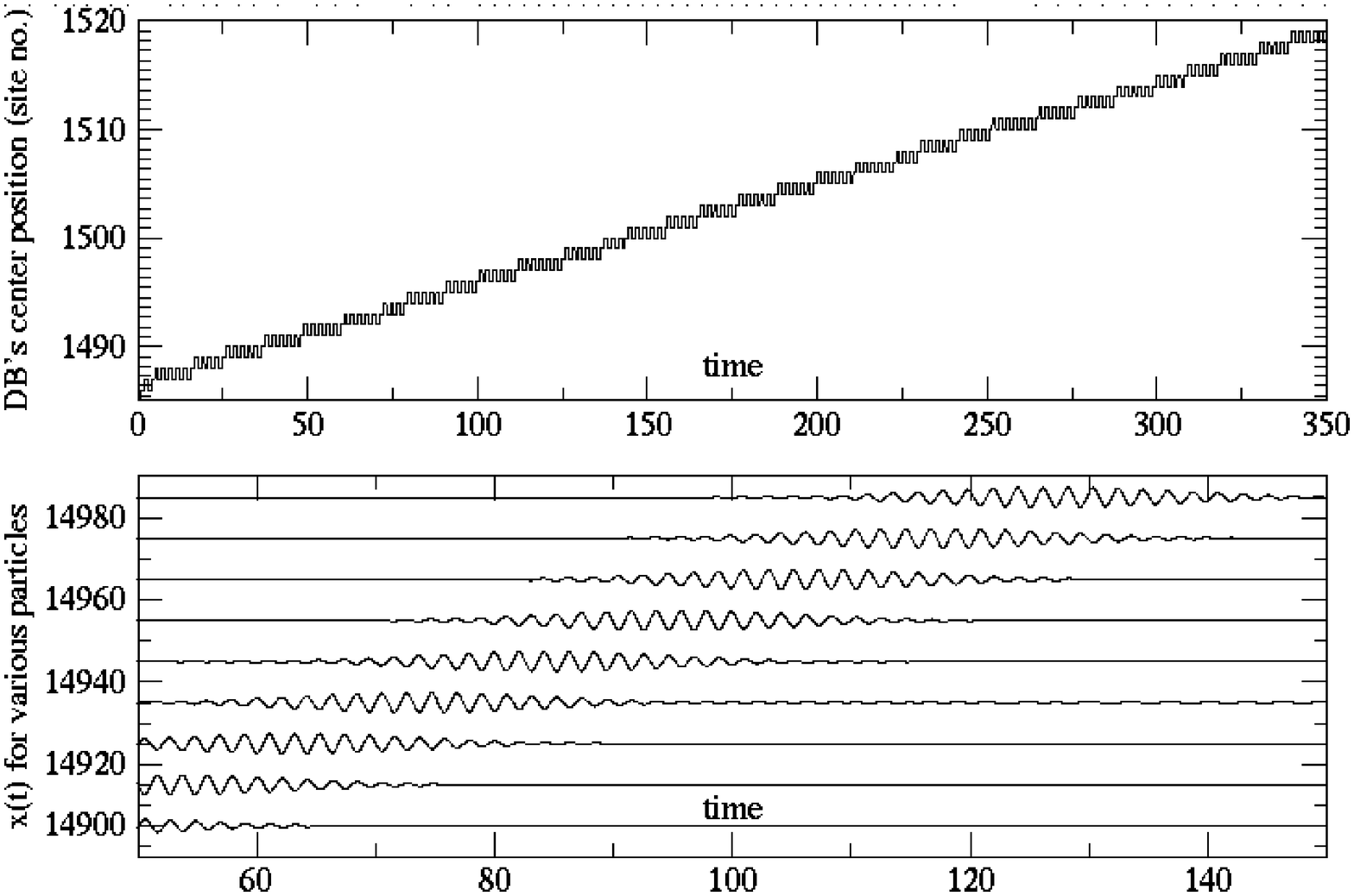,width=8cm}}
\label{figure4a}
\end{minipage}
}\vspace{0.5cm}\\
\subfigure[light-heavy junction, $m_B=1.002$: transmission of the DB]{
\begin{minipage}{0.5\textwidth}
{
\psfig{figure=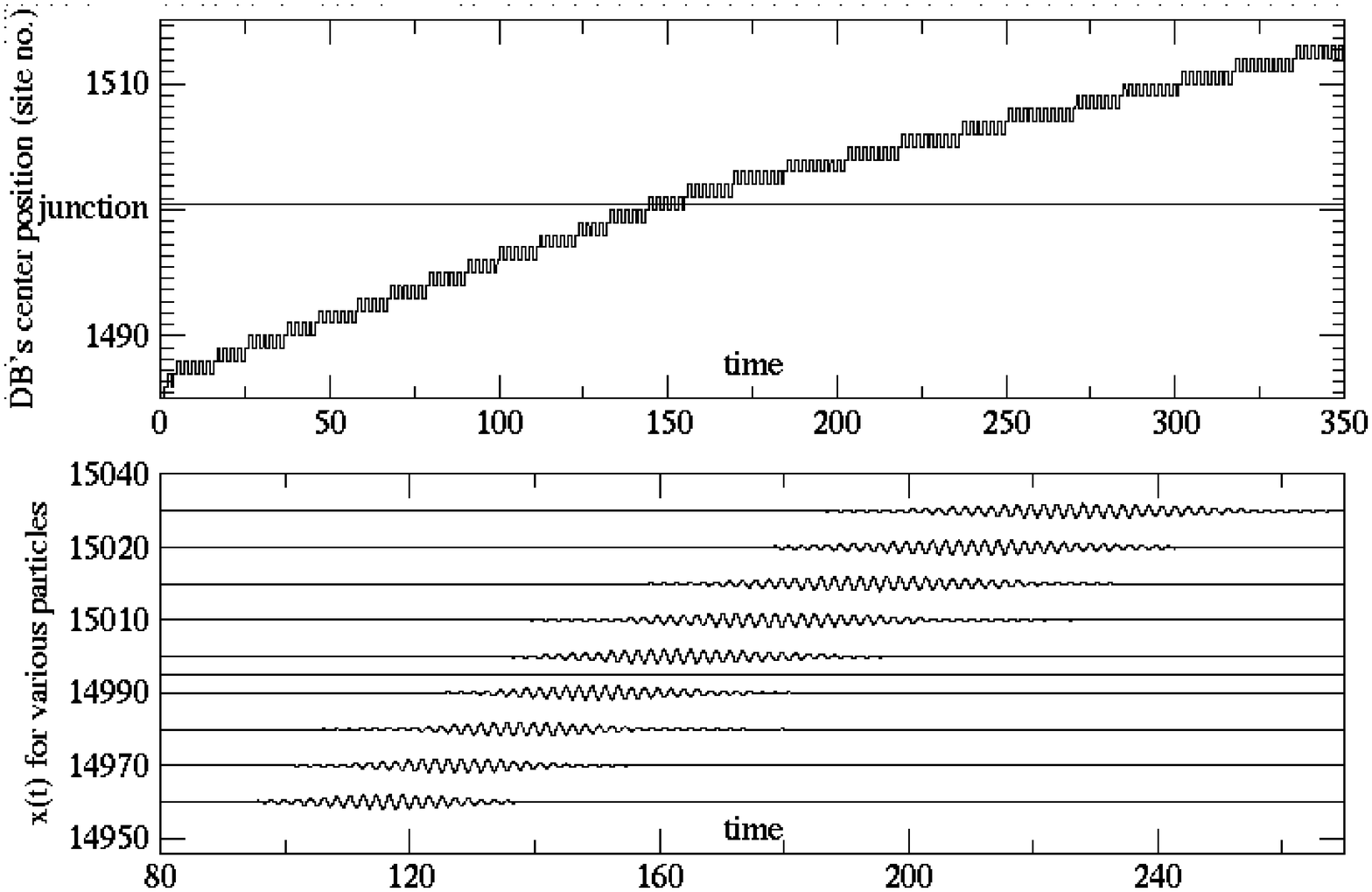,width=8cm}}
\label{figure4b}
\end{minipage}
}\vspace{0.5cm}\\
\subfigure[light-heavy junction, $m_B=1.04$: reflection of the DB]{
\begin{minipage}{0.5\textwidth}
{
\psfig{figure=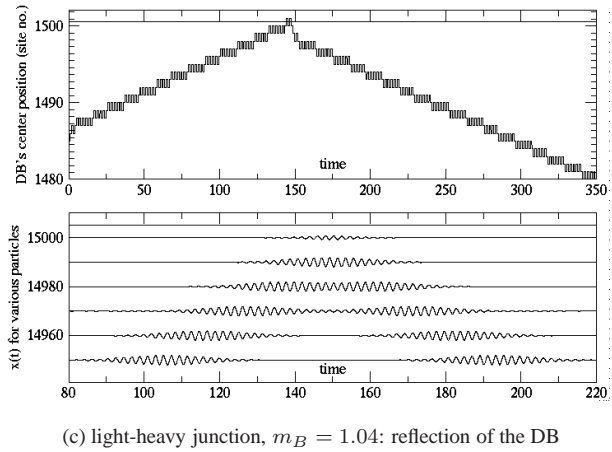,width=8cm}}
\label{figure4c}
\end{minipage}
}
\caption{
Propagation, transmission and reflection of the DB for the 
three situations in Fig. \ref{figure1}.  The associated motion of particles 
in the
region of the DB is also shown.  $T_{DB}=2.1$.}
\label{figure4}
\end{figure}

We present below two comparative sets of pictures of the cases when 
a given DB (a) moves in a homogeneous chain, (b) passes through a 
light-heavy junction, and (c) is reflected at the 
light-heavy junction.  Fig.~\ref{figure1} shows the temporal evolution 
of the configurations of the DB in these three situations, while 
Fig.~\ref{figure4} shows the movement of DB's center along the chain, and 
also the temporal evolution of the elongations of various particles
affected by the DB.

\subsection{The heavy-light junction}
\label{heavylight}
 
As already mentioned above, given that the PNB decreases at a junction 
with $m_B < m_A$ , one can naively predict that the DB will always 
penetrate and continue to move in region B without any hinderance.  
Numerical simulations show that this is indeed the case--at least as 
long as the difference between $m_B$ and $m_A$ is sufficiently small.  
For example, in the particular case of 
$m_B=0.99$ (recall that in simulations we took $m_A=1$) 
we studied the dependence of the characteristics of the 
`transmitted' DB on those of the `incident' one. First of all, one 
notices that the transmitted breather takes some time to ``adjust" 
to the new environment (and `heavier' DBs take a longer time to 
adjust than the `lighter' ones). During this period, the DB loses 
energy and adjusts its final energy to the smaller mass of region B.

The transmitted DB (within estimated errors) has the same period as the
incident one (i.e., the adjustment is such that it preserves DB's 
frequency).  After this transient period, the DB reaches a constant 
`asymptotic' velocity. In general, there is no simple relationship 
between this asymptotic velocity $v_B$ and the characteristics of the 
incident DB, namely its initial velocity (in region A) $v_A$ and 
its period $T_{DB}$.

\begin{figure}
{
{\psfig{figure=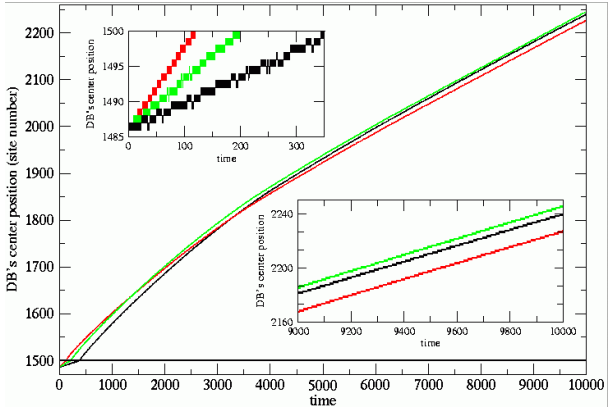,width=8cm}}}
\caption{A heavy--light ($m_B=0.99$) junction. For various initial
velocities $v_A$ of a DB with a given period ($T_{DB}=2.1$) one gets 
approximately the same asymptotic velocity $v_B$ in region B (see the 
lower inset).  The upper inset depicts the DB's position at very early 
time before reaching the junction, while the lower inset depicts the DB's
asymptotic trajectory.}
\label{figure5}
\end{figure}

However, there is a tendency towards `uniform' velocities after 
transmission through the junction for a given DB.  Namely, for a 
DB with period $T_{DB}$ and different velocities in region A, $v_A$, 
the effect of entering region B is to reduce the dispersion of 
these velocities, i.e., the dispersion of the asymptotic velocities, 
$v_B$, is smaller [\onlinecite{footnote3}]. This 
observation is illustrated in Fig.~\ref{figure5} for a given DB (with  
$T_{DB}=2.1$) and for three representative initial velocities $v_A$ 
(chosen, respectively, as the lower and upper limits of the 
velocities that could be obtained through the ``kicking" method 
described above in Sec. \ref{DBgeneration}, and one value in-between 
these limits).

It was relatively more difficult to investigate the dependence of 
the asymptotic velocity $v_B$ on  DB's period $T_{DB}$, simply 
because it is rather difficult with the kicking method to obtain the 
same velocity for DBs of different frequencies.  However, we managed 
to obtain four DBs of periods varying between $2.2$ and $2.5$ (with 
a step $0.1$) and almost (within $4\%$) the same initial velocity. 
The simulations show no simple monotonic dependence of $v_B$ on $T_{DB}$.

Next, we focus on the most important part of this section (that will 
clarify the meaning of sufficiently small difference between $m_B$ and 
$m_A$).  Specifically, how does the behavior of a given DB depend 
on the value of $m_B$?  To analyze this, we  ran systematic simulations 
for a given DB (we chose one with  $T_{DB}=2.1$ and an initial velocity 
$v_A=0.928$, that corresponds to a kicking coefficient $\lambda=0.7$), 
and for various values of $m_B$.  The behavior of the DB at the 
junction is rather complex, and can be described essentially as 
follows: the DB, entering the region of lower mass, has ``extra" energy. 
During an ``adjusting period" (that might take from about ten to hundred 
DB periods),  this extra energy is redistributed between: (i) the kinetic 
energy of DB's 
translation as a whole (the DB is accelerated upon entering  region B); 
(ii) perturbations 
(which we address later) in the A and also in the B part of the chain; and 
(iii) a slight decrease of transmitted DB's period (i.e., increase of 
its configurational energy).  
The redistribution of energy between these elements is a delicate process, 
and it depends on the mass difference between regions A and B.  When the 
mass difference is sufficiently small, up to, say, $(m_A-m_B)/m_A=0.4$, 
the 
predominant phenomena are (i) and (ii)--the perturbations being 
small-amplitude ones, i.e., phonons that move rapidly far away from the 
junction, in both A and B part.  

\begin{figure}
{
{\psfig{figure=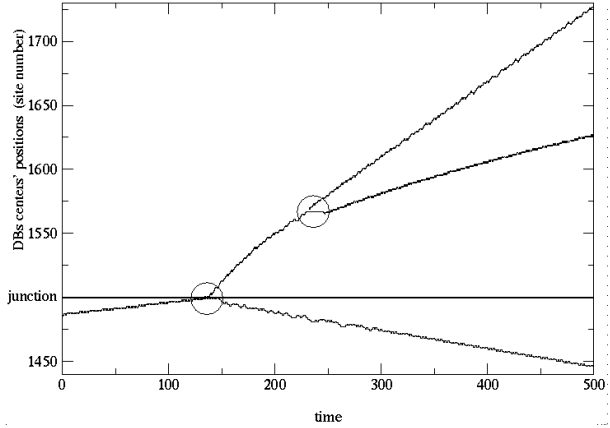,width=8cm}}}
\caption{A heavy--light ($m_B=0.50$) junction.  An initial DB
($T_{DB}=2.1$)  splits into a reflected and a transmitted DB. Later on, the
transmitted DB furhter splits into two other DBs (the circles on the figure
indicate the  regions of the splittings). Note that when the mass difference is
even  larger, the transmitted DB (that has progressively less energy) might
split  into three or even four smaller DBs. Decreasing $m_B$ furhter leads 
practically to the disappearance of the transmitted DB, and to a substantial 
phonon creation.}
\label{figure6}
\end{figure}

\begin{figure}
{
{\psfig{figure=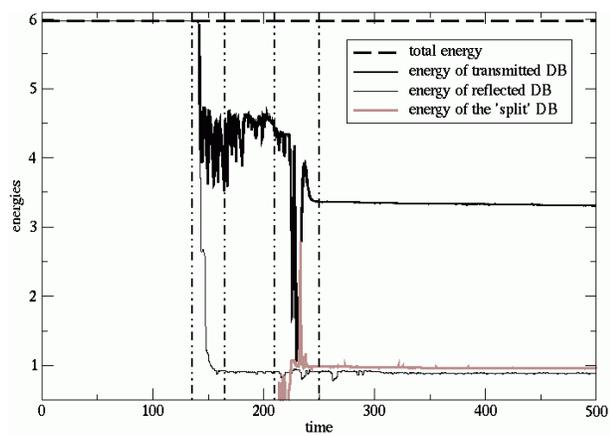,width=8cm}}}
\caption{The energy associated with the phenomena described in Fig.~\ref{figure6}. The dashed-dotted lines on the figure delimit the intervals
of the occurrence of
the splitting
phenomena, when there is no net separation between the resulting DBs, 
i.e., no clear separation of their energies.}
\label{figure7}
\end{figure}

When the mass difference is even larger, we find that in A part there 
are not simply phonons that appear, but a {\it reflected DB}: the initial 
DB, arriving at the junction, is split into a reflected DB and a 
transmitted one. Moreover, the transmitted DB is usually (nonlinearly) 
unstable and subsequently splits into two (or sometimes more) other DBs. 
See Fig.~\ref{figure6} for a realization of these phenomena: the trajectory 
of the initial DB, the reflected one, the transmitted DB and its subsequent 
splitting into two other DBs. Figure~\ref{figure7} offers the energy 
variation associated with these phenomena.  We note that the total  
energies of the resulting DBs never sum up to the initial energy due to
the phonon losses in the chain that accompany all these processes. To our 
knowledge, such DB splitting has not been noticed before.  If we continue 
to decrease $m_B$, the reflected DB (that is initially very weak 
energetically as compared to the transmitted one) becomes progressively 
more energetic, while the transmitted DB becomes progressively weaker and 
finally dissappears in region B leaving only rapidly-moving phonons in 
its wake.  The end product is the ``strong" (i.e., 
large-amplitude) reflected DB in region A.

\section{Interaction of a DB with a `mass ramp'}
\subsection{The Peierls-Nabarro barrier for a `ramp'} 
Consider that in region B the mass of the particles varies slightly,
linearly, as one moves away from the junction point, i.e., the mass of 
the $k$-th particle in B part is 
\begin{equation}
m_B(k)=m_A(1+k\Delta)\,,
\end{equation} 
where $\Delta >0$ corresponds to an up mass-ramp, while $\Delta <0$ 
to a down mass-ramp, and for analytic calculation purposes we consider 
that  $|\Delta|\ll 1$.  A double analytical expansion in $\varepsilon_A$ 
and $\Delta$ allows us to estimate the shape function for the equivalents 
of the odd- and even-type configurations, and therefore to estimate the 
Peierls-Nabarro barrier the DB must overcome in order to move up to site
$k$ in region B.  The barrier (with details given in the Appendix) is 
found to be:
\begin{eqnarray}
\Delta E_{PN}^{r}(k)&=&E_{odd}^{r}(k)-E_{even}^{h(A)}=
\Delta E_{PN}^{h(A)}\nonumber\\
&&\hspace{-1.5cm}+m_A\omega_{DB}^2 
\left(\frac{\alpha_A}{\beta_A}\right)
\left\{\Delta
\left[(0.77906 k + 0.77906)\,\varepsilon_A^{-1}\right.\right.\nonumber\\
&&\left.\left.\hspace{-1.5cm}-(0.16470
k + 0.16470) \right]\right.
\nonumber\\
&&\hspace{-1.5cm}\left.+\Delta^2
\left[(0.38953 k^2+0.77906 k +2.56207)\,
\varepsilon_A^{-1}\right.\right.\nonumber\\
&&\left.\left.\hspace{-1.5cm} +0.91605-3.2585\varepsilon_A\right]\right\}\,,
\label{barrierk}
\end{eqnarray} where the superscript $r$ refers to the ramp.

\subsection{The `up-ramp'}
\label{upramp}

\begin{figure}
\subfigure[]{
\begin{minipage}{0.5\textwidth}
{
\psfig{figure=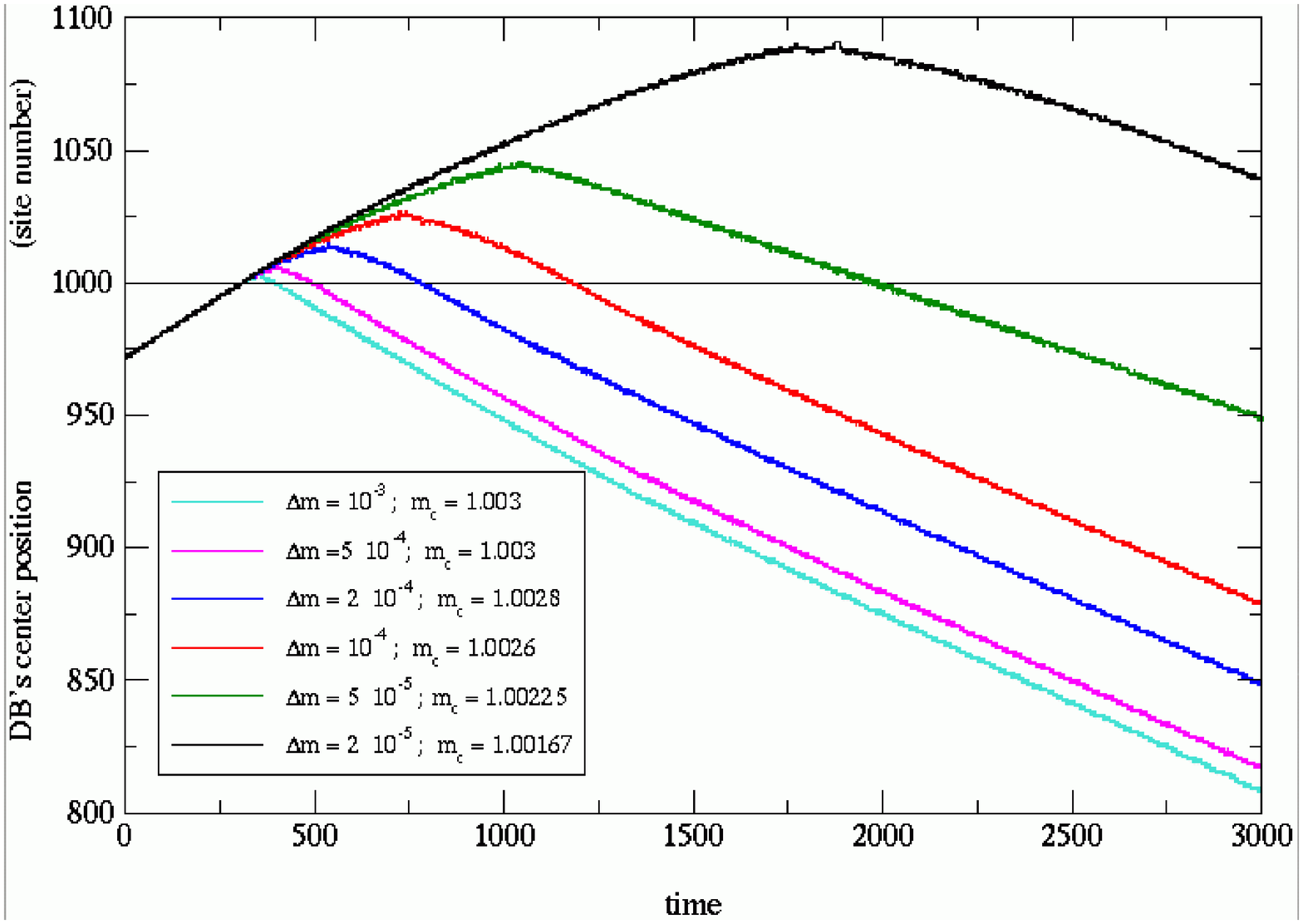,width=8cm}}
\label{figure8a}
\end{minipage}
}\vspace{0.5cm}\\
\subfigure[]{
\begin{minipage}{0.5\textwidth}
{
\psfig{figure=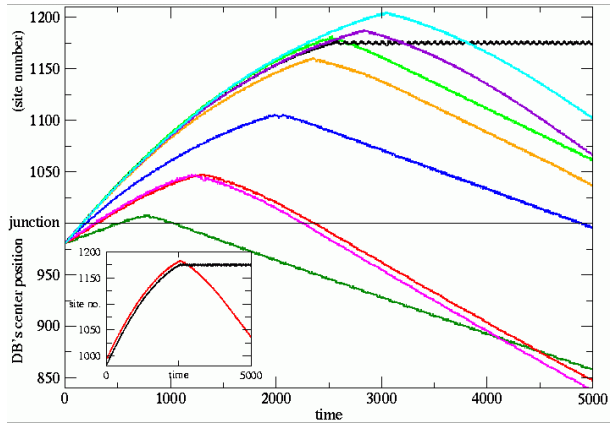,width=8cm}}
\label{figure8b}
\end{minipage}
}\vspace{0.5cm}\\
\subfigure[]{
\begin{minipage}{0.5\textwidth}
{
\psfig{figure=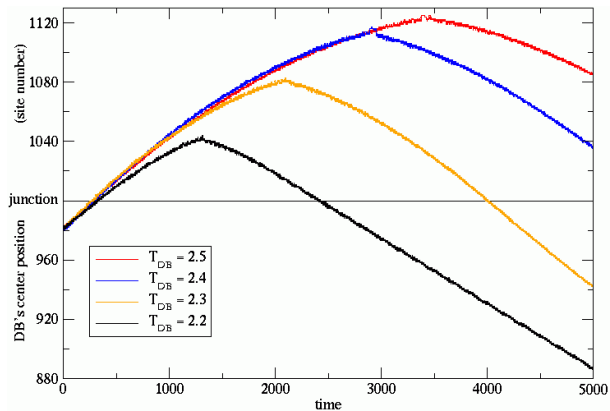,width=8cm}}
\label{figure8c}
\end{minipage}
}
\caption{The behavior of a DB on an up-ramp depending on (a) the slope 
of the ramp ($T_{DB}=2.1$) and (b) its initial velocity $v_A$. Note 
that the DB can also get trapped on the ramp; but, as shown in the 
inset, a slight perturbation--for example, a slight modification of 
the initial conditions--can lead to the disappearance of trapping. 
($T_{DB}=~2.1$).  (c) The behavior of a DB as a function of the 
period $T_{DB}$ of the DB. (Note the limited possibilities to obtain
DBs of various periods with rigorously the same initial velocity $v_A$.)}
\label{figure8}
\end{figure}

This corresponds to the case $\Delta > 0$.
The main result is that a DB that enters 
the B part of the chain is finally reflected (at some point within the 
B chain) and returns to part A. Note that: 

(a) The point where the DB is reflected, i.e., the critical mass on 
the ramp, $m_{crit}^r$, depends on the `slope' $\Delta$ of the ramp 
and is generally different from the value $m_{crit}^j$ that corresponds 
to the case of an abrupt junction, see Sec. \ref{lightheavy}.  This can 
be seen by equating the (critical values of the) most energetic odd-type 
configurations in B chain in the case of an abrupt junction and of a ramp, 
and finding the relationship between $m_{crit}^j$ and $m_{crit}^r$.  In 
the particular case shown in Fig.~\ref{figure8a}, we note that the 
critical mass decreases with decreasing slope of the ramp and that it is
smaller than the value for the junction case.

(b) For a given ramp, the critical mass increases with increasing initial 
velocity of a given DB (with a fixed frequency), see Fig.~\ref{figure8b}.  
Sometimes the DB can get trapped, as seen in the inset of this figure.  
Note, however, that if one changes slightly DB's initial position in  
region A (without changing its initial velocity), then the DB is no 
longer trapped, but reflected, see the inset.  Thus, trapping seems to 
be a rather delicate phenomenon, that depends on `how' (i.e., with what 
precise configuration and relative phase difference between sites) the 
DB arrives at the trapping site. 

(c) For a given ramp, the critical mass seems to increase with 
a decrease in DB's period (i.e., it is larger for `heavier' DBs for 
the same initial velocity), see Fig.~\ref{figure8c}.  Note that in 
all these cases there is a typical temporal evolution of the energy 
of the DB.  Before entering region B, one recognizes the usual small 
energy loss in an uniform FPU chain; then in the ramp part there is a 
somewhat smaller energy loss (presumably the DB is a little bit 
further away from the phonon band) that becomes progressively smaller 
when the DB is decelerated on the ramp.  At a certain moment, the DB 
starts to `descend' the ramp, to increase its velocity, and its energy 
loss increases progressively, again up to the usual loss in the 
homogeneous chain.  When the DB `ascends' the ramp, its configurational 
energy averaged over a  period (and the corresponding Peierls-Nabarro 
barier) increase at the expense of its translational energy.  Therefore, 
at a certain moment, the DB does no longer have sufficient ``extra 
energy" to overcome the barrier and it is reflected (and sometimes it 
may get trapped). Rolling down the hill, it recuperates its translational 
energy and when it gets out from region B and re-enters region A it
has almost the same velocity as its initial one in region A.  The 
transmission and reflection are ``almost elastic", in fact the DB loses 
a little bit less energy than it loses normally during its movement 
in a uniform chain.
    
\subsection{The `down-ramp'} 
\begin{figure}
{
{\psfig{figure=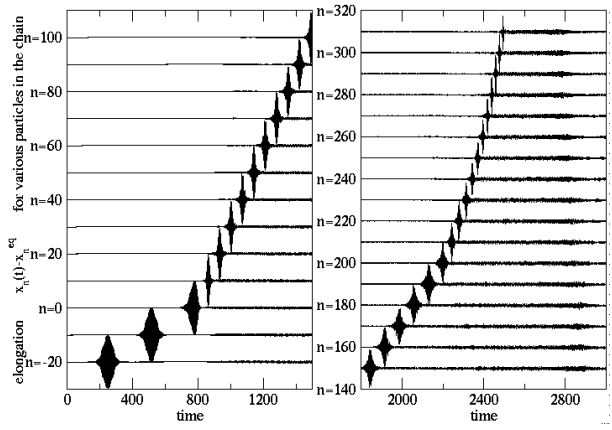,width=8cm}}}
\caption{Acceleration of a DB on a down-ramp.  Site n=0 corresponds to 
the last site in the region before the onset of the ramp. The slope of 
the ramp is $\delta = 0.0018$. On the y axis we approached the sites 
by 9.9 lattice constants ($a=10$ units) 
in order to increase the resolution.}
\label{figure9}
\end{figure}

Consider now that in the region B the mass of the particles decreases 
from one particle to another with the small quantity $m_A\,\Delta < 0$.  
An illustration of DB's typical behavior is given in Fig.~\ref{figure9}.  
As the DB enters the ramp it accelerates with a concomitant narrowing of 
its shape and the emission of some radiation (phonons). This is clearly 
seen as a change in slope (left panel).  At later times we observe another 
change in slope signifying further acceleration of the DB with significant 
radiation and emission of smaller breathers (right panel).  One can also 
observe other secondary, small-energy DBs that may form at later stages. 
Nonetheless, these phenomena are highly complex and beyond our current 
level of understanding. 
     
\section{Conclusions and perspectives} 

We have systematically explored the transport properties of a 
discrete breather in a nonlinear chain comprising two segments 
with differing mass, specifically in an FPU chain.  We considered 
abrupt junctions (light-heavy and heavy-light) as well as (up and 
down) mass ramps.  We studied the trapping, reflection, transmission 
and splitting of the DB as a function of junction type, mass 
difference, breather frequency and velocity.  The DB splitting, 
trapping and reflection may take place either at the junction or 
at a particular particle within the ramp.  We also estimated the 
Peierls-Nabarro barrier for the different cases to understand the 
DB transport across a junction or within a ramp. However, the approach 
for calculating the PNB is based on the fundamental assumption that the
period (frequency) of the DB does not change ``significantly" during 
its movement, which has its limitations, as shown by the simulations 
and indicated above in various cases. Therefore, we can rely on this 
method only at a qualitative level.   
 
In the present paper we exclusively focused on two segments with 
slightly different masses.  It would be interesting to explore a 
junction (or ramp) between two segments with the same mass but 
with differing strength of either the harmonic ($\alpha$) or 
anharmonic ($\beta$) interaction parameter of the FPU chain.  
This is under investigation and our preliminary results do not 
demonstrate a qualitatively different picture compared to the mass 
case.  In addition, if we consider an A--B--A mass sandwich structure 
then there is a distinct possibility that the breather will get 
trapped inside the B segment.  By a suitable choice of the mass 
profile one may envision a `breather lens'.  This is currently explored
and preliminary results agree with these conjectures.  
We believe that our results are {\it not specific to the FPU chain}.  
Other nonlinear potentials should lead to generically similar results. 
Many open questions remain, e.g. better estimates for site-to-site 
traversal time of a DB, influence of the ``secondary" frequencies of 
the DB on its behavior (for example, on 
the envelope of temporal oscillations of 
energy), a better understanding of the nonlinear
instability that leads to DB splitting 
(reflected/transmitted, and afterwards to
the secondary splitting of the transmitted DB), and consequently,
to the complex behavior
on a down-ramp, etc. An experimental realization of our findings in 
low-dimensional electron-phonon coupled materials [\onlinecite{agrawal}], 
e.g. conjugated polymers [\onlinecite{kress,avn2}] and metal-halogen 
chains [\onlinecite{avn3}], using different {\it isotopes} would be 
quite instructive in unraveling the interesting transport properties of 
breathers with potential applications.   

%Summary of main findings.  Mention different lenghth scales, e.g.
%lattice parameter, characteristic length of the breather.  Also,
%different time scales such as breather period, phonon period, etc.   

\acknowledgments{} We are indebted to M. Iba\~{n}es and G. P. Tsironis 
for insightful discussions and help with the numerics.  This work 
has been supported  by the European Union under the RTN project
LOCNET (HPRN-CT-1999-00163), by the U.S. Department of Energy, and 
by the Direcci\'on General de Ense\~nanza Superior e Investigaci\'on
Cient\'{\i}fica (Spain) under Project no. BFM2000-0624.   
A.S. gratefully acknowledges a fellowship from Iberdrola (Spain).     

\section{Appendix: PNB for various configurations}
In this appendix we present the relevant details of estimating the 
Peierls-Nabarro barrier for the different cases discussed in the text. 
\subsection{The homogeneous chain}
\subsubsection{The odd-type mode}
\label{oddparitymode}

It is characterized by $\xi_n = \xi_{-n} \equiv (-1)^|n| \eta_n$ ($\eta_n$, 
the reduced shape function, being positive for all $n$), together with
the condition $\zeta_0=\eta_0=1$ (that gives the normalization of the shape
function). The equations for the reduced shape function read, respectively:
\begin{eqnarray}
&&{\Lambda \varepsilon}=\frac{2 - \varepsilon (1+\eta_1)}{
(1+\eta_1)^3}\,\,, \mbox {for}\,\,\, n=0\,,
\nonumber\\
\eta_n&=&\frac{\varepsilon}{4}(2\eta_n+\eta_{n+1}+\eta_{n-1})+
\frac{\Lambda \varepsilon}{4}[(\eta_n+\eta_{n+1})^3 \nonumber \\ 
&+&(\eta_n+\eta_{n-1})^3]
\,\,, \mbox {for}\,\,\, n \geq 1\,.\nonumber\\
\label{etaodd}
\end{eqnarray}
Here 
\begin{equation}
\Lambda=\frac{3 \beta A_{odd}^2}{4 \alpha}\,.
\label{Lambdaodd}
\end{equation}
Note the singular behavior in $\varepsilon^{-1}$ of the square of the amplitude,
$A_{odd}^2$ (e.g., this means that 
the `heavier' the DB, the larger its amplitude).
Using the series expansions in $\varepsilon$ for $\xi_n$, Eq. 
(\ref{seriesxin}), and the corresponding ones for the reduced shape 
function in Eq. (\ref{etaodd}), and ordering the corresponding powers 
of $\varepsilon$, one can show that the series $\eta_n^{(j)}$ (for a 
fixed $j$, i.e., for a fixed order in the perturbative expansion in 
$\varepsilon$) rapidly converges to zero with increasing $n$; more 
rapidly for the case of small $j$s than for larger ones 
[\onlinecite{footnote4}].  Finally, one is led to the following 
expressions for particles' shape function:
\begin{eqnarray}
&&\xi_0=1\,,\nonumber\\
&&\xi_1=\xi_{-1}=(-1)[0.52305 + 0.15113\, \varepsilon + 0.08549\,
\varepsilon^2 \nonumber\\&&
\hspace{1cm}+ {\cal O}(\varepsilon^3)]\,,\nonumber\\ 
&&\xi_2=\xi_{-2}= 0.02305+0.15691 \,\varepsilon+0.12643\,\varepsilon^2
\nonumber\\
&&\hspace{1cm}+ {\cal O}(\varepsilon^3)\,,\nonumber\\
&&\xi_3=\xi_{-3}=(-1)[0.00580\, \varepsilon + 0.04239\,
\varepsilon^2\nonumber\\
&&\hspace{1cm}+{\cal O}(\varepsilon^3)]\,,\nonumber\\
&&\xi_n = {\cal O}(10^{-6})\,\,,\,|n| \geq 4\,\,,
\label{oddmode}
\end{eqnarray} 
together with the dependence of the amplitude $A_{odd}$ on DB's frequency, mass
$m$ of the particles 
(through $\varepsilon$), $\alpha$, and $\beta$:
\begin{equation}
\Lambda=\frac{3 \beta A_{odd}^2}{4
\alpha}={0.56609}\,{\varepsilon}^{-1}-0.59960+ 0.02366 \,\varepsilon +
{\cal O}(\varepsilon^2)\,.
\end{equation}
All these lead finally to the following expression for the configurational
energy of the odd-parity mode:
\begin{eqnarray}
E_{odd}^{h}&=&m \omega_{DB}^2 \left(\frac{\alpha}{\beta}\right) 
[{0.38953}\, {\varepsilon}^{-1}-0.16470 \nonumber \\ 
&-& 0.12386 \,\varepsilon + {\cal O}(\varepsilon^2)]\,\,.
\label{energyodd}
\end{eqnarray}

\subsubsection{The even-type mode}
\label{evenparitymode}

It is characterized by the presence of two `main peaks', $\xi_0=-\xi_1=1$, and
also by a staggered shape:
$\xi_n=-\xi_{-n+1}=(-1)^{|n|} \eta_n$, with the reduced positive shape function
$\eta_n$.
In this case, the equations for the reduced shape function read, respectively:
\begin{eqnarray}
&&{\Lambda \varepsilon}=\frac{4 - \varepsilon (3+\eta_1)}{8+
(1+\eta_1)^3}\,\,, \mbox {for}\,\,\, n=0,1\,,
\nonumber\\
&&\eta_n=\frac{\varepsilon}{4}(2\eta_n+\eta_{n+1}+\eta_{n-1})+
\frac{\Lambda
\varepsilon}{4}[(\eta_n+\eta_{n+1})^3\nonumber\\
&&\hspace{1cm}+(\eta_n+\eta_{n-1})^3]
\,\,, \mbox {for}\,\,\, n \geq 2\,.\nonumber\\
\label{etaeven}
\end{eqnarray}
Here 
\begin{equation}
\Lambda=\frac{3 \beta A_{even}^2}{4 \alpha}\,\,.
\label{Lambdaeven}
\end{equation}
Finally, one is led to the following expressions for particles' shape function:
\begin{eqnarray}
&&\xi_0=-\xi_1=1\,,\nonumber\\
&&\xi_2=-\xi_{-1}=0.16579 + 0.31767\,\varepsilon + 0.13806\, \varepsilon^2
+ {\cal O}(\varepsilon^3)\,,\nonumber\\ 
&&\xi_3=-\xi_{-2}=(-1)[0.00048+0.04438
\,\varepsilon+0.10766\,\varepsilon^2 \nonumber\\
&&\hspace{2cm}+
{\cal O}(\varepsilon^3)]\,,\nonumber\\ 
&&\xi_4=-\xi_{-3}= 0.00012 \,\varepsilon + 0.01115\,
\varepsilon^2\nonumber\\
&&\hspace{2cm}+{\cal O}(\varepsilon^3)\,,\nonumber\\
&&\xi_n = {\cal O}(10^{-6})\,\,,\,n \geq 5, \,n \leq -4\,,
\label{evenmode}
\end{eqnarray} 
and the equation for the amplitude:
\begin{equation}
\Lambda=\frac{3 \beta A_{even}^2}{4 \alpha}=\frac{0.41735}{\varepsilon}-0.38670+
0.02077 \varepsilon + {\cal O}(\varepsilon^2)\,\,.
\end{equation}
The corresponding configurational energy:
\begin{eqnarray}
E_{even}^{h}&=&m \omega_{DB}^2
\left(\frac{\alpha}{\beta}\right)
[{0.38117}\,{\varepsilon}^{-1}-0.15705\nonumber\\
&-& 0.10559\,
\varepsilon + {\cal O}(\varepsilon^2)]\,.
\end{eqnarray}
Note that 
$E_{even}^{h} < E_{odd}^{h}$, i.e., as already noticed 
[\onlinecite{Sandusky1992}], {\it the
even-type mode is more stable than the odd-type one}.

\subsection{The junction}
\begin{figure}
{
{\psfig{figure=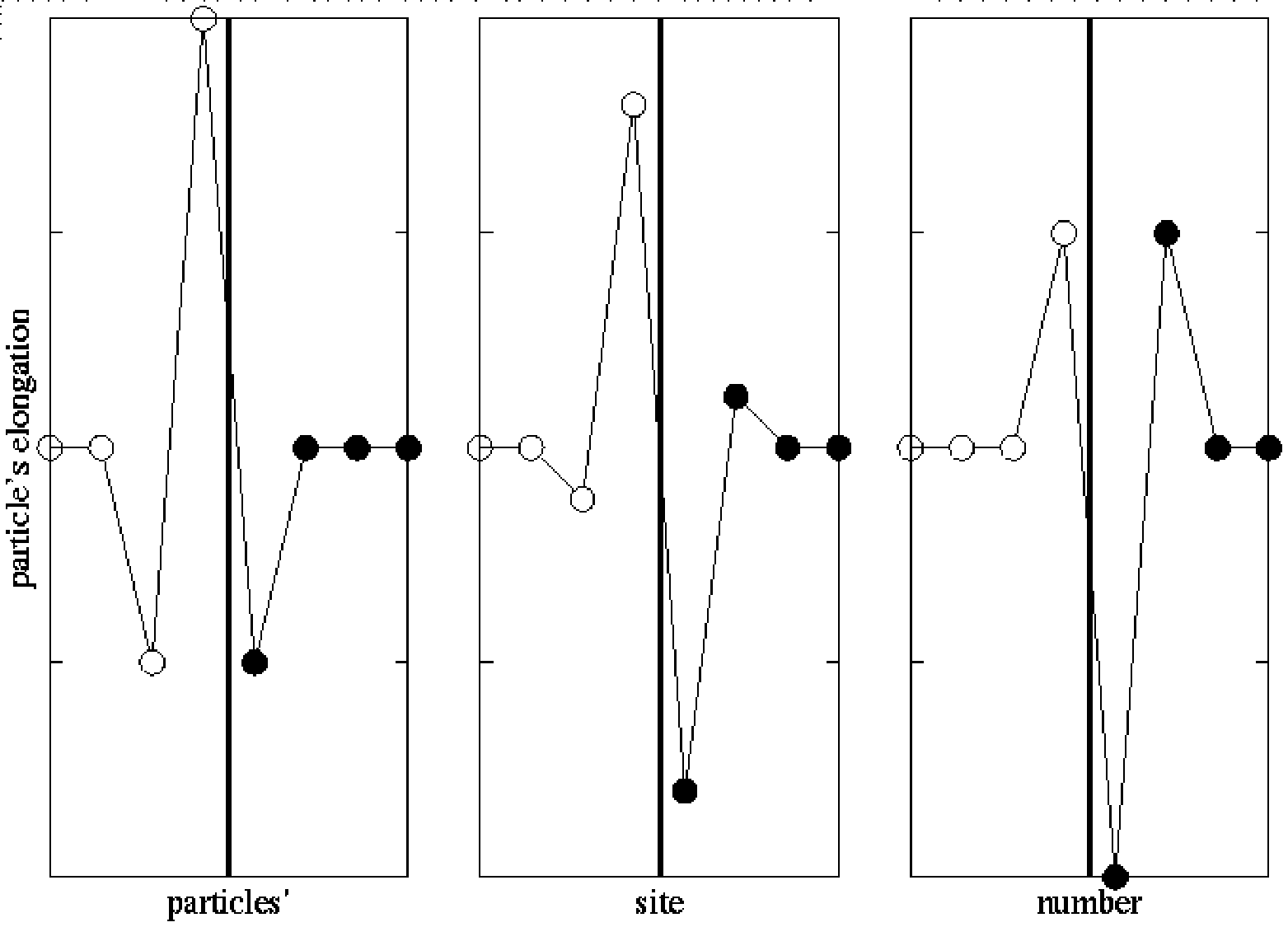,width=8cm}}}
\caption{Schematic representation of the succession of the odd- and 
even- type of configurations for a DB traversing a junction from left to
right. The white and black circles correspond, respectively,
to particles in the A and B parts of the chain.}
\label{figure10}
\end{figure}

We refer to Fig.~\ref{figure10} to follow the different configurations of 
the DB moving from left to right through the junction. As indicated in the 
text, Eq. (\ref{doubleseries}), we used a double perturbation expansion 
of the envelope function--in both $\varepsilon_A$ (evaluated with respect 
to the parameters of the A chain, see the text) and $\delta$ to compute 
the different configurations. Note that the convergence in $\delta$ is
not as good as that for $\varepsilon_A$; namely, $\delta$ should be 
$10^{2}$ times {\it less} than $\varepsilon_A$ in order to get the same 
degree of correction for the same order of expansion in $\delta$ as in 
$\varepsilon_A$. However, the details of the calculations are lenghty and, 
because they present no conceptual difficulty, not  given here. 
Instead, we  give the expressions for the configurational energies of 
the DB in its successive appearances--as these  allow us to compute 
the various Peierls-Nabarro barriers it encounters.
\subsubsection{Configuration I}
It is of the odd-type--it corresponds to the first panel in Fig.~\ref{figure10}: the site of maximum elongation is in part A of the chain.
Its energy is found to be:
\begin{eqnarray}
E_{odd}^{j(I)}&=&E_{odd}^{h(A)}+m_A\omega_{DB}^2\left(
\frac{\alpha_A}{\beta_A}\right)\,\nonumber\\
&&\hspace{-2cm}\times\left[\delta
\left(0.13793\, \varepsilon_A^{-1}-0.01573 + 0.0056\, \varepsilon_A\right)\right.
\nonumber\\
&&\hspace{-2cm}\left.+\delta^2
\left(0.49933 \,\varepsilon_A^{-1}+0.21657 -0.69346\, \varepsilon_A\right)\right].
\label{eoddjunctionI}
\end{eqnarray}
Correspondingly, the first barrier that the DB has to overcome is 
the one between an even-type 
configuration in the homogeneous A chain and this
configuration, namely:
\begin{eqnarray}
E_{odd}^{j(I)}-E_{even}^{h(A)}&=&\Delta
E_{PN}^{h(A)}+ m_A\omega_{DB}^2
\left(\frac{\alpha_A}{\beta_A}\right)\nonumber\\ 
&&\hspace{-2.5cm}\times\left[\delta
\left(0.13793 \,\varepsilon_A^{-1}-0.01573 + 0.0056 \,\varepsilon_A\right)\right.
\nonumber\\
&&\hspace{-2.5cm}\left.+\delta^2
\left(0.49933 \,\varepsilon_A^{-1}+0.21657 -0.69346\, \varepsilon_A\right)\right]\,.
\label{barrierI}
\end{eqnarray}

\subsubsection{Configuration II}
It is of an even-type and corresponds to the second panel of Fig.~\ref{figure10}: there are two sites with large elongations, one of them 
in part A of the chain, the other one in part B of the chain.
The corresponding energy, up to ${\cal O}(\varepsilon_A^2,\delta^2)$, is: 
\begin{eqnarray}
E_{even}^{j(II)}&=&E_{even}^{h(A)}+m_A\omega_{DB}^2
\left(\frac{\alpha_A}{\beta_A}\right)\nonumber\\
&&\hspace{-1.5cm}\times\left[\delta
\left(0.38117\, \varepsilon_A^{-1}-0.07852 \right)\right.
\nonumber\\
&&\hspace{-1.5cm}\left.+\delta^2
\left(-0.83444 \,\varepsilon_A^{-1}-0.32507 +1.11672\, \varepsilon_A\right)\right].
\label{eevenjunction}
\end{eqnarray}
Therefore, the energy difference between configurations II and I is:
\begin{eqnarray}
E_{even}^{j(II)}-E_{odd}^{j(I)}&=&-\Delta E_{PN}^{h(A)}+
m_A\omega_{DB}^2\left(\frac{\alpha_A}{\beta_A}\right)\nonumber\\
&&\hspace{-2.5cm}\times\left[\delta
\left(-0.24324\, \varepsilon_A^{-1}-0.06280 - 0.00526 \,\varepsilon_A\right)\right.
\nonumber\\
&&\hspace{-2.5cm}\left.+\delta^2
\left(-1.33278 \,\varepsilon_A^{-1}-0.54163+1.81108 \,\varepsilon_A\right)\right]\,.
\end{eqnarray}

\subsubsection{Configuration III}
Again of the odd-type, it corresponds to the third panel in Fig.~\ref{figure10}: the site of maximum elongation is now in region B of the chain.
Its energy:
\begin{eqnarray}
E_{odd}^{j(III)}&=&E_{odd}^{h(A)}+m_A\omega_{DB}^2
\left(\frac{\alpha_A}{\beta_A}\right)\nonumber\\
&&\hspace{-1.5cm}\times\left[\delta
\left(0.64113 \,\varepsilon_A^{-1}-0.14880 -0.00526 \,\varepsilon_A\right)\right.
\nonumber\\
&&\hspace{-1.5cm}\left.+\delta^2
\left(0.75093\, \varepsilon_A^{-1}+0.21657 -0.68820 \,
\varepsilon_A\right)\right]\,.
\label{eoddjunctionII}
\end{eqnarray}
Therefore, the energy barrier between configuration II and configuration III is:
\begin{eqnarray}
E_{odd}^{j(III)}-E_{even}^{j(II)}&=&\Delta
E_{PN}^{h(A)}+ m_A\omega_{DB}^2
\left(\frac{\alpha_A}{\beta_A}\right)\nonumber\\
&&\hspace{-2.5cm}\times\left[\delta
\left(0.25996 \,\varepsilon_A^{-1}-0.07045 -0.00526\, 
\varepsilon_A\right)\right.
\nonumber\\
&&\hspace{-2.5cm}\left.+\delta^2
\left(1.58537 \,\varepsilon_A^{-1}+0.54163 -1.80493 \,
\varepsilon_A\right)\right]\,. 
\label{barrierIII}
\end{eqnarray}
After this, the DB is essentially in the homogeneous B part [of mass 
$m_B=m_A(1+\delta)$]; the energies of the even- and odd-type 
configurations in B are:
\begin{eqnarray}
E_{even}^{h(B)}&=&E_{even}^{h(A)}+m_A\omega_{DB}^2
\left(\frac{\alpha_A}{\beta_A}\right)\nonumber\\
&&\hspace{-1.5cm}\times\left[\delta
\left(0.76234 \,\varepsilon_A^{-1}-0.15705\right)\right.
\nonumber\\
&&\hspace{-1cm}\left.+\delta^2
\left(0.38117\, \varepsilon_A^{-1}\right)\right]\,,
\label{eevenB}
\end{eqnarray}
\begin{eqnarray}
E_{odd}^{h(B)}&=&E_{odd}^{h(A)}+m_A\omega_{DB}^2
\left(\frac{\alpha_A}{\beta_A}\right)\nonumber\\
&&\hspace{-1.5cm}\times\left[\delta
\left(0.77906 \,\varepsilon_A^{-1}-0.16470\right)\right.
\nonumber\\
&&\hspace{-1cm}\left.+\delta^2
\left(0.38953\, \varepsilon_A^{-1}\right)\right]\,.
\label{eoddB}
\end{eqnarray}
Therefore, on one side, the energy difference between the 
configuration III and
the even configuration of chain B is:
\begin{eqnarray}
E_{even}^{h(B)}-E_{odd}^{j(II)}&=&-\Delta
E_{PN}^{h(A)}+ m_A\omega_{DB}^2
\left(\frac{\alpha_A}{\beta_A}\right)\nonumber\\
&&\hspace{-2.5cm}\times\left[\delta
\left(0.12122\, \varepsilon_A^{-1}-0.00808 +0.00526 \,\varepsilon_A\right)\right.
\nonumber\\
&&\hspace{-2.5cm}\left.+\delta^2
\left(-0.36976 \,\varepsilon_A^{-1}-0.21657 +0.68820
\,\varepsilon_A\right)\right]\,,
\end{eqnarray}
and, on the other hand, the PNB in the homogeneous B part is:
\begin{eqnarray}
E_{odd}^{h(B)}-E_{even}^{h(B)}&=&\Delta
E_{PN}^{h(A)}+ m_A
\omega_{DB}^2\left(\frac{\alpha_A}{\beta_A}\right)\nonumber\\  
&&\hspace{-3cm}\times[\delta (0.01672\,
\varepsilon_A^{-1}-0.00765)+\delta^2
(0.00836\,\varepsilon_A^{-1})]\,. 
\label{PNBhB}
\end{eqnarray}

\subsection{The ramp}
Consider that the ramp has a ``slope" $\Delta$, i.e., 
the mass of the $k$-th site in the ramp is 
\begin{equation}
m_B(k)=m_A(1+k \Delta)\,\,.
\end{equation}
Note that $\Delta >0$ corresponds to an up-ramp, while $\Delta <0$ to a
down-ramp.
Consider first a configuration of the even-type 
where the
two sites with maximum elongation are $k-1$ and $k$.
The corresponding configurational energy is found to be:
\begin{eqnarray}
E_{even}^{r}(k)&=&E_{even}^{h(A)}+m_A\omega_{DB}^2
\left(\frac{\alpha_A}{\beta_A}\right)\nonumber\\
&&\hspace{-2cm}\times\left\{\Delta
\left[(0.76234 k - 0.38117)\, \varepsilon_A^{-1}\right.\right.\nonumber\\
&&\left.\left.\hspace{-2cm}+(-0.15705
k + 0.07852)  \right]\right.
\nonumber\\
&&\hspace{-2cm}\left.+\Delta^2
\left[(0.38117 k^2-0.38117 k -1.35992)
\,\varepsilon_A^{-1}\right.\right.\nonumber\\
&&\left.\left.\hspace{-2cm} -0.85452 + 3.28732 \,\varepsilon_A\right]\right\}\,.
\label{eevenramp}
\end{eqnarray}
Consider then the next configurational step in the displacement of the DB from
left to right on the ramp, i.e., an odd-parity type of configuration
that is centered on the
$k$-th site, i.e., the $k$-th site is the 
one that has the maximum elongation.
The energy of this configuration is:
\begin{eqnarray}
E_{odd}^{r}(k)&=&E_{odd}^{h(A)}+m_A\omega_{DB}^2
\left(\frac{\alpha_A}{\beta_A}\right)\nonumber\\
&&\hspace{-2cm}\times\left\{\Delta
\left[(0.77906 k + 0.77906)\,\varepsilon_A^{-1}\right.\right.\nonumber\\
&&\left.\left.\hspace{-2cm}-(0.16470
k + 0.16470)  \right]\right.
\nonumber\\
&&\hspace{-2cm}\left.+\Delta^2
\left[(0.38953 k^2+0.77906 k +2.56207)\,
\varepsilon_A^{-1}\right.\right.\nonumber\\
&&\left.\left.\hspace{-2cm} +0.91605-3.25850\,\varepsilon_A\right]\right\}\,.
\label{eoddramp}
\end{eqnarray}

Correspondingly, the energy barrier to overcome while moving 
from site $k-1$ to
site $k$ is:
\begin{eqnarray}
E_{odd}^{r}(k)-E_{even}^{r}(k)&=&
\Delta E_{PN}^{h(A)}+m_A\omega_{DB}^2 
\left(\frac{\alpha_A}{\beta_A}\right)\nonumber\\
&&\hspace{-3cm}\times\left\{\Delta
\left[(0.01672 k + 1.16023)\, \varepsilon_A^{-1}\right.\right.\nonumber\\
&&\hspace{-3cm}\left.\left.-(0.00765
k + 0.24322)  \right]\right.
\nonumber\\
&&\hspace{-3cm}\left.+\Delta^2
\left[(0.00836 k^2+ 1.16023 k +3.92199)
\varepsilon_A^{-1}\right.\right.\nonumber\\
&&\hspace{-3cm}\left.\left. +1.77057- 6.54584\,\varepsilon_A\right]\right\}\,\,.
\label{EPNramp}
\end{eqnarray}
At the dominant order in both $\Delta$ and $\varepsilon_A$, one
finds an increase in the PNB for an up mass-ramp ($\Delta >0$), and 
a decrease of the barrier for a down mass-ramp ($\Delta <0$).

%%%%%%%%%FIGURE 1%%%%%%%%%%%%%%%%%%%%%%%%%%%%%%%%%%%%%%%%%%%%%%%%%%%%%%%%%%%

%%%%%%%%%%%%%%%%%%FIGURE 2%%%%%%%%%%%%%%%%%%%%%%%%%%%%%%%%%%%%%%%%%%%%%%%%%%

%%%%%%%%%%%%%%%%%%%%%%%%%%%%%%%%%%%%%%%%%%%%%%%%%%%%%%%%%%%%%%%%%%%%%%%%%%%%%

%%%%%%%%%%%%%%%%%FIGURE 3 %%%%%%%%%%%%%%%%%%%%%%%%%%%%%%%%%%%%%%%%%

%%%%%%%%%%%%%%%%%%%%%%%%%%%%%%%%%%%%%%%%%%%%%%%%%%%%%%%%%%%%%%%%%%%%%%%%%%%

%%%%%%%%%%%%%%%%%%%%%%%%%%%%%%%%%%%%%%%%%%%%%%%%%%%%%%%%%%%%%%%%%%%%%
%%%%%%%%%%%%%%FIGURE 4%%%%%%%%%%%%%%%%%%%%%%%%%%%%%%%%%%%%%%%%%%%%%%%%
%%%%%%%%%%%%%%%%%%%%%%%%%%%%%%%%%%%%%%%%%%%%%%%%%%%%%%%%%%%%%%%%%%%%%%%%%

%%%%%%%%%%%%%%%%%%FIGURE 5%%%%%%%%%%%%%%%%%%%%%%%%%%%%%%%%%%%%%%%%%%%%%%%%%%%

%%%%%%%%%%%%%%%%%%%%%%%%%%%%%%%%%%%%%%%%%%%%%%%%%%%%%%%%%%%%%%%%%%%%%%%%%%%%%%
%%%%%%%%%%%%%%%%%%FIGURE 6%%%%%%%%%%%%%%%%%%%%%%%%%%%%%%%%%%%%%%%%%%%%%%%%%%%

%%%%%%%%%%%%%%%%%%%%%%%%%%%%%%%%%%%%%%%%%%%%%%%%%%%%%%%%%%%%%%%%%%%%%%%%%%%%%%
%%%%%%%%%%%%%%%%%%FIGURE 7 %%%%%%%%%%%%%%%%%%%%%%%%%%%%%%%%%%%%%%%%%%%%%%%%%%%

%%%%%%%%%%%%%%%%%%%%%%%%%%%%%%%%%%%%%%%%%%%%%%%%%%%%%%%%%%%%%%%%%%%%%%%%%%%%%%
%%%%%%%%%%%%%%FIGURE 8%%%%%%%%%%%%%%%%%%%%%%%%%%%%%%%%%%%%%%%%%%%%%%

%%%%%%%%%%%%%%%%%%%%%%%%%%%%%%%%%%%%%%%%%%%%%%%%%%%%%%%%%%%%%
%%%%%%%%%%%%%%FIGURE 9%%%%%%%%%%%%%%%%%%%%%%%%%%%%%%%%%%

%%%%%%%%%%%%%%%%%%%%%%%%%%%%%%%%%%%%%%%%%%%%%%%%%%%%%%%%%%%%
%%%%%%%%%%%%%%%FIGURE 10 - APPENDIX %%%%%%%%%%%%%%%%%%%%%%%

%%%%%%%%%%%%%%%%%%%%%%%%%%%%%%%%%%%%%%%%%%%%%%%%%%%%%%%%%%


\begin{thebibliography}{99}
\bibitem{footnote3}{Note that an effect that is reminiscent 
of this one was reported in [\onlinecite{Ibanes2002}]. Namely, the 
authors considered a moving DB that traverses a curved part of a modified 
FPU chain (i.e., a chain with both first- and second-neighbor 
interactions of the FPU type).  It was found that the velocity of 
the DB after leaving the curved region {\it was (almost) the same}, 
regardless of the velocity of the DB before entering the curved 
region. Note also that, as the authors emphasize, the same result 
was obtained when the ``geometrical perturbation" (i.e., the 
presence of the curved part) of the chain was replaced by an 
initial random perturbation in the transverse DB direction.}
\bibitem{footnote4}{Recall 
also that $\sum_n \xi_n =0$ (the center of mass conservation), which 
leads to $\sum_n (-1)^n \eta_n ^{(j)}=0$, $j \geq 0$.  This renders 
such a behavior of the coefficients of the perturbative series 
intuitively more accessible.}
\end{thebibliography}
\end{document}